\documentclass[twocolumn]{aastex631}

\usepackage{mathrsfs }
\usepackage{natbib}
\usepackage{amssymb}
\usepackage{apjfonts}
\usepackage{ulem} 
\usepackage{graphicx}
\begin{document}
\clearpage

\title{JWST/NIRSpec Measurements of the Relationships Between Nebular Emission-line Ratios and Stellar Mass at $\lowercase{z}\sim 3-6$}

\author[0000-0003-3509-4855]{Alice E. Shapley}\affiliation{Department of Physics \& Astronomy, University of California, Los Angeles, 430 Portola Plaza, Los Angeles, CA 90095, USA}
\email{aes@astro.ucla.edu}

\author[0000-0001-9687-4973]{Naveen A. Reddy}\affiliation{Department of Physics \& Astronomy, University of California, Riverside, 900 University Avenue, Riverside, CA 92521, USA}

\author[0000-0003-4792-9119]{Ryan L. Sanders}\altaffiliation{NHFP Hubble Fellow}\affiliation{Department of Physics and Astronomy, University of California, Davis, One Shields Ave, Davis, CA 95616, USA}

\author[0000-0001-8426-1141]{Michael W. Topping}\affiliation{Steward Observatory, University of Arizona, 933 N Cherry Avenue, Tucson, AZ 85721, USA}

\author[0000-0003-2680-005X]{Gabriel B. Brammer}\affiliation{Cosmic Dawn Center (DAWN), Denmark}\affiliation{Niels Bohr Institute, University of Copenhagen, Lyngbyvej 2, DK2100 Copenhagen \O, Denmark}

\shortauthors{Shapley et al.}

\shorttitle{Emission-line Ratios at $z\sim 3-6$}

\begin{abstract}
We analyze the rest-optical emission-line ratios of star-forming galaxies at $2.7\leq z<6.5$ drawn from the Cosmic Evolution Early Release Science (CEERS) Survey, and their relationships with stellar mass (${\rm M}_*$). Our analysis includes both line ratios based on 
the [NII]$\lambda 6583$ feature  -- [NII]$\lambda6583$/H$\alpha$, ([OIII]$\lambda5007$/H$\beta$)/([NII]$\lambda6583$/H$\alpha$) (O3N2), and [NII]$\lambda6583$/[OII]$\lambda3727$ -- and those those featuring $\alpha$ elements -- [OIII]$\lambda5007$/H$\beta$, [OIII]$\lambda5007$/[OII]$\lambda3727$ (O$_{32}$), ([OIII]$\lambda\lambda4959,5007$+[OII]$\lambda3727$)/H$\beta$ (R$_{23}$), and 
[NeIII]$\lambda3869$/[OII]$\lambda3727$. Given the typical flux levels of  [NII]$\lambda6583$ and  [NeIII]$\lambda3869$, which are undetected in the majority of individual CEERS galaxies at $2.7\leq z<6.5$, we construct composite spectra in bins of ${\rm M}_*$ and redshift.  Using these composite spectra, we compare the relationships between emission-line ratios and ${\rm M}_*$ at $2.7\leq z<6.5$ with those observed at lower redshift. 
While there is significant evolution towards higher excitation (e.g., higher [OIII]$\lambda5007$/H$\beta$, O$_{32}$, O3N2), and weaker nitrogen emission (e.g., lower [NII]$\lambda6583$/H$\alpha$ and [NII]$\lambda6583$/[OII]$\lambda3727$) between $z\sim 0$ and $z\sim 3$, we find in most cases that there is no significant evolution in the relationship between line ratio and ${\rm M}_*$ {\it beyond} $z\sim 3$.  
The [NeIII]$\lambda3869$/[OII]$\lambda3727$ ratio is anomalous in showing evidence for significant elevation at $4.0\leq z<6.5$ at fixed mass, relative to $z\sim 3.3$. Collectively, however, our empirical results suggest no significant evolution in the mass-metallicity relationship at $2.7\leq z<6.5$.  Representative galaxy samples and metallicity calibrations based on existing and upcoming {\it JWST}/NIRSpec observations will be required to translate these empirical scaling relations into ones tracing chemical enrichment and gas cycling, 
and distinguish among descriptions of feedback in galaxy-formation simulations at $z>3$

\end{abstract}

\section{Introduction}
\label{sec:intro}
The rest-frame optical nebular emission-line spectrum of star-forming galaxies is rich 
with information probing their gas, heavy elements, dust, and massive stars. The pattern of rest-optical emission
lines is also known to vary systematically as a function of stellar mass, (${\rm M}_*$).
This variation occurs primarily as a result of the dependence of galaxy metallicity on galaxy stellar mass.
This so-called ``mass-metallicity relation" (MZR) has been traced using vast samples in the      
local universe drawn from the Sloan Digital Sky Survey \citep[SDSS; e.g.,][]{tremonti2004,andrews2013}, and,
with smaller, yet still statistical, power all the way out to $z\sim 3$ 
\citep[e.g.,][]{onodera2016,kashino2017,topping2021,sanders2021}. A secondary, additional dependence of
metallicity on star-formation rate (SFR) has been discovered as well
\citep[e.g.,][]{ellison2008,mannucci2010}, the ``Fundamental Metallicity Relation," (FMR) which appears not to evolve
significantly between $z\sim 0$ and $z\sim 3$ \citep{sanders2021}.

Previously restricted to $z\lesssim3.5$, studies of the rest-optical emission-line properties of star-forming
galaxies can now be extended deep into the reionization epoch with {\it JWST}. Even the very first 
NIRSpec spectroscopic data released to the public revealed the promise of {\it JWST} for estimating the gas-phase
chemical abundances of star-forming galaxies at $z\sim 7.5-8.5$ \citep[e.g.,][]{arellano2022,curti2022}. Subsequent
remarkable NIRSpec spectra have revealed the rest-optical nebular properties of individual galaxies at $z=9.5$ \citep{williams2022},
and, now, $z=10.6$ \citep{bunker2023}. The low metallicities of the $z>8$ targets observed by {\it JWST}, 
given their stellar masses and SFRs, suggest that they do not fall on the FMR that describes galaxies at $z\leq 3$ \citep{curti2022,williams2022,langeroodi2022}.
However, it is challenging to draw broad conclusions from such a small sample,
which may or may not be representative of the full population at these redshifts, and questions remain about which calibrations can be used to accurately infer metallicity at such high redshifts. 

Larger samples are clearly needed to place detailed, single-object results in context. The CEERS Early Release
Science program \citep[][; Finkelstein et al., in prep.]{finkelstein2022a,finkelstein2022b} provides one of the first such opportunities \citep[see also][]{matthee2022,mascia2023,cameron2023}, including 6
NIRSpec pointings in the EGS field using the medium resolution ($R\sim1000$) grating, and 6 pointings
using the $R\sim100$ prism, targeting collectively $\sim 1000$ galaxies with photometric redshifts over the range
$z\sim 0.5-12$ \citep{fujimoto2023}. Already, CEERS is providing a window into the excitation properties
of star-forming galaxies out to $z\sim 9$ \citep{sanders2023a,tang2023}. Even $z\sim 3.5-6.5$ represents uncharted territory for rest-optical spectroscopic emission-line studies of star-forming galaxies. Furthermore,
{\it JWST}/NIRSpec provides access to a broader complement of rest-optical emission lines at $z\sim 3.5-6.5$ than those covered for
the most distant galaxies with {\it JWST} spectra \citep{shapley2023}. Based on its richness, with simultaneous coverage of [OII]$\lambda3727$, [NeIII]$\lambda3869$,
H$\beta$, [OIII]$\lambda\lambda4959,5007$, H$\alpha$, and [NII]$\lambda6583$ in CEERS/NIRSpec medium-resolution spectra, we focus in this work
on the redshift range $z=2.7-6.5$. Robust stellar mass estimates for galaxies in this redshift range enable us to analyze the relationships
among rest-optical emission-line ratios and stellar masses. Tracing these relationships for both $\alpha$-elements (e.g., O, Ne)
and nitrogen is a first crucial step towards establishing metallicity scaling relationships at $z>3$, which also requires a robust
calibration between emission-line properties and oxygen abundance \citep{bian2018,sanders2020,sanders2021,nakajima2023}.

In \S\ref{sec:obs}, we describe
observations and samples analyzed here. In \S\ref{sec:results}, we present results on the observed
relationships between rest-optical emission-line ratios and stellar mass, 
measured continuously from $z\sim 3-6$. 
In \S\ref{sec:discussion}, we compare with other recent work and consider the implications of these new measurements for the evolution of metallicity scaling relations among galaxies.
Throughout, we adopt cosmological parameters of
$H_0=70\mbox{ km  s}^{-1}\mbox{ Mpc}^{-1}$, $\Omega_{\rm m}=0.3$, and
$\Omega_{\Lambda}=0.7$, and a \citet{chabrier2003} IMF.

\section{Observations, Sample, and Measurements} 
\label{sec:obs}

\subsection{Observations}
\label{sec:obsobs}
Our analysis is based on the public  NIRSpec Micro-Shutter Assembly (MSA) data from the CEERS program
\citep[Program ID:1345;][; Arrabal Haro et al., in prep.]{finkelstein2022a, finkelstein2022b}.  We analyze
data from 6 medium-resolution ($R\sim1000$) NIRSpec pointings in the 
AEGIS field, spanning $1-5\mu$m with the  grating/filter combination of G140M/F100LP, G235M/F170LP, and G395M/F290LP. These 6 pointings included a total
sample of 318 distinct targets, observed for 3107 seconds in each grating/filter combination. Two-dimensional (2D) data processing steps,
one-dimensional (1D) extraction, slit-loss corrections, band-to-band flux calibration, and the measurement of emission-line fluxes are all described in detail in \citet{shapley2023}, \citet{sanders2023a}, and \citet{reddy2023}.

\begin{figure}[t!]
\centering
\includegraphics[width=1.0\linewidth]{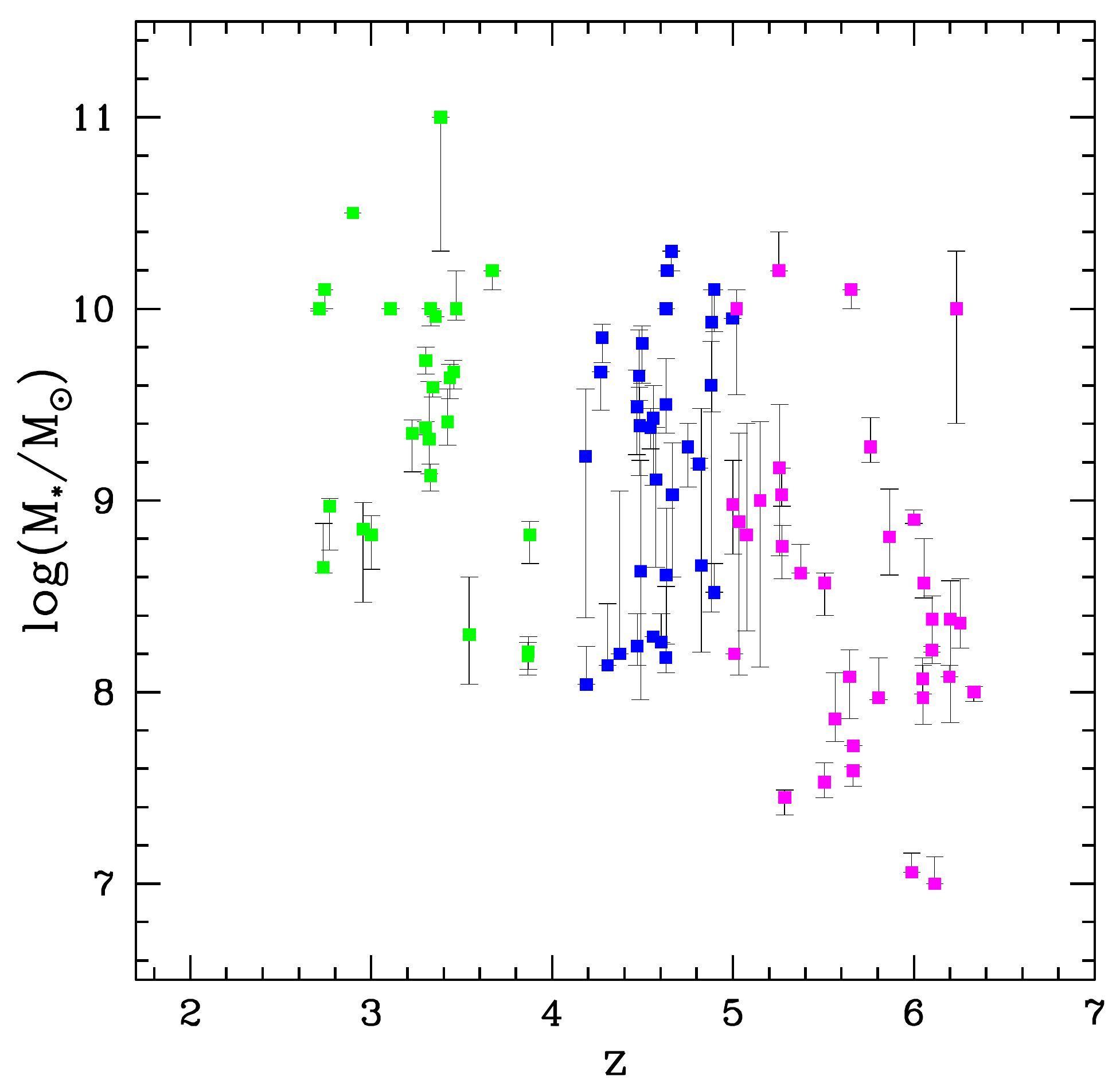}
\caption{Stellar mass vs. redshift for all 94 CEERS galaxies at $2.7 \leq z \leq 6.5$ analyzed in this work. Galaxies at $z=2.7-4.0$ are color-coded green; those at $z=4.0-5.0$ are shown in blue; finally, those at $z=5.0-6.5$ are indicated in magenta.
}
\label{fig:sample}
\end{figure}

We extracted spectra for 252 CEERS targets, and measured spectroscopic redshifts for 231. As described in \citet{shapley2023}, we modeled existing {\it HST}/ACS and WFC3, {\it JWST}/NIRCam, {\it Spitzer}/IRAC, and ground-based photometry for CEERS galaxy targets using the FAST program \citep{kriek2009}, assuming the stellar population synthesis models of \citet{conroy2009}. We adopted delayed-$\tau$ star-formation histories, where $SFR(t)\propto t \times exp(-t/\tau)$, and $t$ is the time since the onset of star formation. Robust spectral energy distributions (SEDs), corrected for emission-line fluxes, were determined for 210 of the 231 galaxies with spectroscopic redshifts, for which we accordingly obtained stellar mass estimates. 

\begin{figure*}[t!]
\centering
\includegraphics[width=0.8\linewidth]{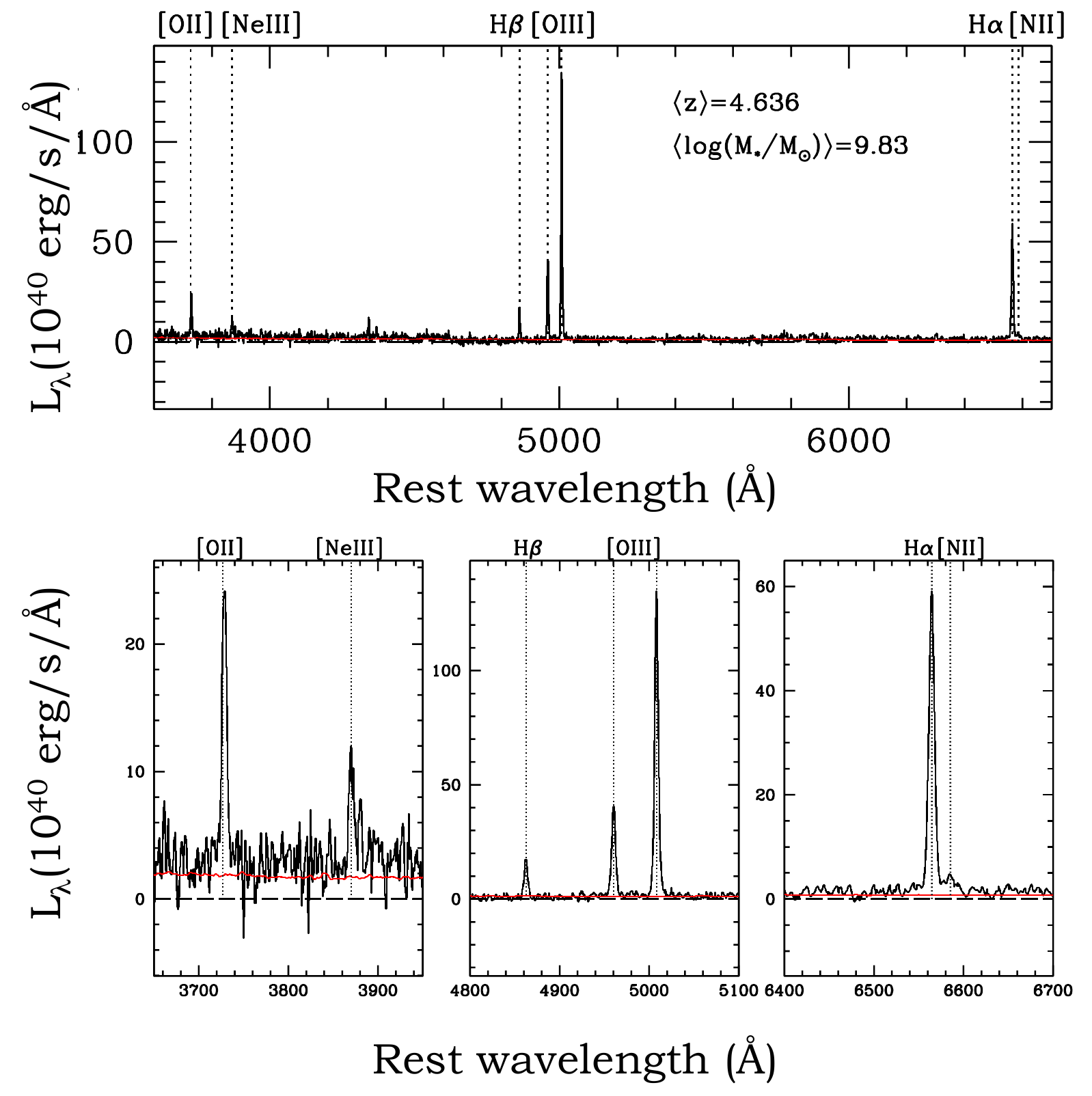}
\caption{Composite spectrum of the high-mass bin of CEERS galaxies at $z=4.0-5.0$. The composite spectrum is indicated with a black curve, while the error spectrum is shown in red. In the top row, the full composite spectrum is shown, along with the mean redshift and stellar mass. The bottom row includes three zoomed-in regions covering the vicinity of the key nebular lines analyzed in this work. These emission lines are labeled and marked with vertical dotted lines in both full and zoomed-in spectral panels. 
}
\label{fig:spec}
\end{figure*}

\subsection{Sample}
\label{sec:obssamp}
We restricted the current analysis to CEERS galaxies at $2.7 \leq z < 6.5$,
which spans in redshift from the current ground-based high-redshift limit for H$\alpha$ measurements, up to the limit at which H$\alpha$ and [NII]$\lambda6583$ can be measured with the NIRSpec G395M/F290LP set-up. We also require an estimate of stellar mass, and a lack of spectroscopic indication of active galactic nucleus (AGN) activity, such as $\log({\rm [NII]}\lambda6583/{\rm H}\alpha) \geq -0.3$ or broad H$\alpha$ emission. These criteria yield a sample of 94 galaxies. In order to search for redshift evolution, we construct three redshift subsamples at $2.7 \leq z < 4$ (27 galaxies),  $4.0 \leq z < 5.0$ (32 galaxies), and $5.0 \leq z < 6.5$ (35 galaxies). In Figure~\ref{fig:sample}, we plot the stellar masses and redshifts for our sample, color-coded according to redshift. In \citet{shapley2023}, we showed that the $2.7 \leq z < 4.0$  and $4.0 \leq z < 5.0$ CEERS samples are representative of main sequence star-forming galaxies \citep{speagle2014}, whereas the $5.0 \leq z < 6.5$ may represent galaxies with higher than average specific SFRs.

\begin{table*}
 \centering
 \caption{Emission-line Properties of CEERS Composite Spectra
 }\label{tab:stacks}
 \setlength{\tabcolsep}{4.5pt}
 \renewcommand{\arraystretch}{1.5}
 \begin{tabular}{rllllllll}
    \hline\hline
   $\log{\left(\frac{\text{M}_*}{\text{M}_\odot}\right)}$$^a$ &
 $N$$^b$ &
$\log\left(\frac{\text{[NII]}}{\text{H}\alpha}\right)$ &
$\log\left(\text{O3N2}\right)$ &
$\log\left(\frac{\text{[NII]}}{\text{[OII]}}\right)$ &
$\log\left(\frac{\text{[OIII]}}{\text{H}\beta}\right)$ &
$\log\left(\text{O}_{32}\right)$ &
$\log\left(\text{R}_{23}\right)$ &
$\log\left(\frac{\text{[NeIII]}}{\text{[OII]}}\right)$  \\
   \hline\hline
   \multicolumn{9}{c}{$z=2.7-4.0$ stacks in bins of ${\rm M}_*$} \\
   \hline
$8.93\pm0.13$  & 12 & $-1.48^{+0.07}_{-0.08}$ & $2.26^{+0.07}_{-0.08}$  & $-1.28^{+0.08}_{-0.09}$ & $0.78^{+0.01}_{-0.01}$ & $0.54^{+0.02}_{-0.03}$ & $0.99^{+0.01}_{-0.02}$ & $-0.41^{+0.03}_{-0.03}$ \\
$10.09\pm0.10$  & 13 & $-0.88^{+0.01}_{-0.01}$ & $1.36^{+0.02}_{-0.02}$  & $-0.90^{+0.03}_{-0.03}$ & $0.48^{+0.01}_{-0.01}$ & $0.01^{+0.02}_{-0.02}$ & $0.84^{+0.02}_{-0.02}$ & $-1.25^{+0.10}_{-0.12}$ \\

   \hline\hline
   \multicolumn{9}{c}{$z=4.0-5.0$ stacks in bins of ${\rm M}_*$} \\
   \hline
$8.76\pm0.15$  & 13 & $-1.46^{+0.08}_{-0.09}$ & $2.24^{+0.08}_{-0.10}$  & $-1.30^{+0.09}_{-0.11}$ & $0.78^{+0.03}_{-0.03}$ & $0.48^{+0.04}_{-0.04}$ & $0.99^{+0.03}_{-0.04}$ & $-0.38^{+0.04}_{-0.04}$ \\
$9.83\pm0.08$  & 13 & $-1.23^{+0.03}_{-0.03}$ & $2.04^{+0.03}_{-0.03}$  & $-1.17^{+0.04}_{-0.05}$ & $0.81^{+0.01}_{-0.02}$ & $0.40^{+0.02}_{-0.02}$ & $1.02^{+0.02}_{-0.02}$ & $-0.55^{+0.05}_{-0.05}$ \\
   \hline\hline
   \multicolumn{9}{c}{$z=5.0-6.5$ stacks in bins of ${\rm M}_*$} \\
   \hline
   $7.98\pm0.11$  & 16 & $-1.36^{+0.09}_{-0.11}$ & $2.15^{+0.09}_{-0.11}$  & $-1.23^{+0.10}_{-0.13}$ & $0.78^{+0.02}_{-0.02}$ & $0.45^{+0.06}_{-0.07}$ & $0.99^{+0.02}_{-0.02}$ & $-0.53^{+0.14}_{-0.20}$ \\
$9.25\pm0.15$  & 15 & $-1.31^{+0.05}_{-0.06}$ & $1.99^{+0.05}_{-0.06}$  & $-1.07^{+0.06}_{-0.07}$ & $0.69^{+0.01}_{-0.02}$ & $0.47^{+0.02}_{-0.02}$ & $0.90^{+0.02}_{-0.02}$ & $-0.49^{+0.06}_{-0.07}$ \\
   \hline\hline
 \end{tabular}
 \begin{flushleft}
 $^{a}$ {Median stellar mass of galaxies in each bin.}
 $^{b}$ {Number of galaxies in each bin. The composite sample includes a total of 82 galaxies with H$\alpha$ detections.}
 \end{flushleft}
\end{table*}

\subsection{Measurements}
\label{sec:obsmeas}
Based on the emission-line measurements from CEERS galaxy spectra, we estimated several line ratios for each galaxy, which form the basis of our analysis. These line ratios include:

$\bullet$ [NII]$\lambda6583$/H$\alpha$ 

$\bullet$ ([OIII]$\lambda5007$/H$\beta$)/([NII]$\lambda6583$/H$\alpha$) (hereafter O3N2)

$\bullet$ [NII]$\lambda6583$/[OII]$\lambda3727$ 

$\bullet$ [OIII]$\lambda5007$/H$\beta$

$\bullet$ [OIII]$\lambda5007$/[OII]$\lambda3727$ (hereafter O$_{32}$)

$\bullet$ ([OIII]$\lambda5007$+[OII]$\lambda3727$)/H$\beta$ (hereafter R$_{23}$)

$\bullet$ [NeIII]$\lambda3869$/[OII]$\lambda3727$

Ratios between lines with small wavelength separation (i.e., [NII]$\lambda6583$/H$\alpha$, O3N2, [OIII]$\lambda5007$/H$\beta$, [NeIII]$\lambda3869$/[OII]$\lambda3727$) do not require corrections for dust attenuation, and are presented as is, while [NII]$\lambda6583$/[OII]$\lambda3727$, O$_{32}$, and R$_{23}$ are corrected for dust attenuation based on the H$\alpha$/H$\beta$ Balmer decrement, assuming an intrinsic H$\alpha$/H$\beta$ ratio of 2.79 and the \citet{cardelli1989} dust law, which has been shown to be appropriate for high-redshift star-forming galaxies \citep{reddy2020}.  

While the majority of our sample has individual detections of [OIII]$\lambda5007$/H$\beta$, O$_{32}$, and R$_{23}$, any line ratio including either [NII] or [NeIII] results in a majority of limits in either one or more required emission lines. In order to utilize the full sample, including limits, and obtain results that are representative of the relationships between emission-line ratios and stellar mass, we constructed composite spectra in two roughly equal-sized bins of stellar mass for each of the three redshift bins. The 82 (out of 94) CEERS galaxies at $2.7 \leq z < 6.5$ with detections of H$\alpha$ were included in the sample for stacking. H$\alpha$ detections were required since individual spectra were normalized to a common H$\alpha$ luminosity before averaging. Emission line fluxes, ratios, and uncertainties were measured from each composite spectrum \citep{reddy2023}. Figure~\ref{fig:spec} shows an example of a CEERS composite spectrum, representing the high-mass bin within the redshift range $z=4.0-5.0$. All key emission lines used in our analysis are labeled.

\section{Results}
\label{sec:results}

We present the relations between emission-line ratios and stellar mass in each of the three CEERS redshift bins spanning $2.7 \leq z < 6.5$. We plot individual CEERS galaxies when a detection or meaningful $3\sigma$ upper- or lower-limit can be derived. Given the low detection rate for some line ratios, we also plot values measured from composite spectra in two bins of stellar mass for each redshift bin. These values are presented in Table~1. For comparison, we indicate the analogous relations for lower-redshift samples. Specifically, in all panels, we plot the distribution of SDSS $z\sim 0$ star-forming galaxies as a grey, 2D histogram \citep{abazajian2009}. We also plot measurements from stacked spectra of star-forming galaxies drawn from the MOSDEF survey.
These include $z\sim 1.5$ measurements from \citet{topping2021}, and $z\sim 2.3$ and $z\sim 3.3$ measurements from \citet{sanders2021}. The $z\sim 2.3$ measurements span [OII]$\lambda3727$, [NeIII]$\lambda3869$, H$\beta$, [OIII]$\lambda5007$, H$\alpha$, and [NII]$\lambda6583$, whereas the $z\sim 1.5$ measurements lack the bluest features ([OII]$\lambda3727$, [NeIII]$\lambda3869$); conversely, the $z\sim 3.3$ measurements lack the reddest features (H$\alpha$ and [NII]$\lambda6583$).

\begin{figure*}[t!]
\centering
\includegraphics[width=1.0\linewidth]{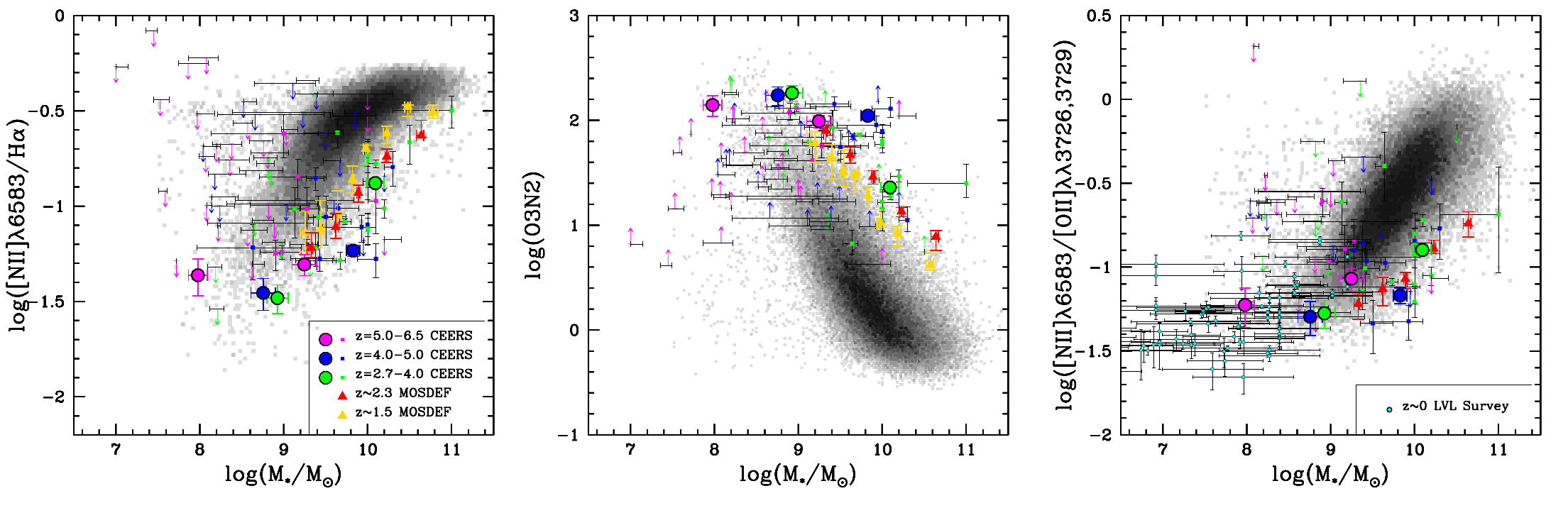}
\caption{Nitrogen-related emission-line ratios vs. stellar mass for CEERS galaxies. These panels show the line ratios [NII]$\lambda6583$/H$\alpha$ (left), O3N2 (center), and [NII]$\lambda6583$/[OII]$\lambda3727$ (right) versus ${\rm M}_*$. All of these ratios include the [NII]$\lambda6583$ feature commonly used to probe ISM conditions at $z\leq 2.5$.  In each panel, the background grayscale histogram corresponds to the distribution of local SDSS star-forming galaxies. Individual CEERS galaxies are color-coded as in Figure~\ref{fig:sample}, and indicated as small squares for detections or downward-facing (upward-facing) arrows for upper (lower) limits. Measurements from stacked CEERS spectra are indicated as large circles with the same color-coding as individual measurements. Measurements from stacked spectra of galaxies in the MOSDEF survey at $z\sim 1.5$ \citep{topping2021} and $z\sim 2.3$ \citep{sanders2021} are shown, respectively, with large gold and red triangles. In the right panel ([NII]$\lambda6583$/[OII]$\lambda3727$), we overplot low-mass galaxies from the Local Volume Legacy (LVL) Survey \citep{berg2012} with small cyan symbols, in order to indicate the regime of primary nitrogen enrichment in which N/O is independent of metallicity.
}
\label{fig:n2plots}
\end{figure*}

\subsection{Line Ratios Including Nitrogen}
\label{sec:results-n2}
We begin with scaling relations including the [NII]$\lambda6583$ line: [NII]$\lambda6583$/H$\alpha$, O3N2, and  [NII]$\lambda6583$/[OII]$\lambda3727$. One basic result concerns the relative faintness of [NII]$\lambda6583$. Out of 82 galaxies with coverage of both [NII]$\lambda6583$ and H$\alpha$ at $z=2.7-6.5$, only 31 have $3\sigma$ detections of [NII]$\lambda6583$. The CEERS composite spectra show that the typical ratio of [NII]$\lambda6583$/H$\alpha$ in this redshift range is $0.04-0.06$, except at $\log({\rm M}_*/M_{\odot})\geq 10$, where it rises to $0.10-0.15$ in the lowest-redshift bin. Indeed, based on both individual datapoints and stacked measurements, there is evidence in Figure~\ref{fig:n2plots} (left) for a scaling between [NII]$\lambda6583$/H$\alpha$ and ${\rm M}_*$ within the CEERS $z=2.7-4.0$ bin, which is consistent with the scaling observed by \citet{sanders2021} for $z\sim2.3$ star-forming galaxies in the MOSDEF survey. On the other hand, MOSDEF star-forming galaxies at $z\sim 1.5$ show a progression towards higher [NII]$\lambda6583$/H$\alpha$ at fixed stellar mass, which continues to $z\sim 0$. At $z\geq 4.0$, no strong trends are recovered between [NII]$\lambda6583$/H$\alpha$ and stellar mass. In addition, the stellar mass range probed at $z>5.0$ extends towards significantly lower values ($\log({\rm M}_*/M_{\odot})\sim 8.0$). At this stellar mass, given the existence of the MZR, nitrogen production falls within the primary
regime, as discussed below \citep[and see Figure~14 of][]{andrews2013}.

O3N2 (Figure~\ref{fig:n2plots}, center) and [NII]$\lambda6583$/[OII]$\lambda3727$ (Figure~\ref{fig:n2plots}, right) tell a similar story, with the CEERS $z=2.7-4.0$ sample overlapping the MOSDEF $z\sim 2.3$ scaling relations, and no strong trends within the CEERS $z>4.0$ galaxies. One possible reason for the lack of strong scaling between [NII]$\lambda6583$/[OII]$\lambda3727$ and mass at $z>4.0$ is that we are probing the low-metallicity regime of primary nitrogen production, where the N/O ratio is independent of metallicity \citep[e.g.,][]{pilyugin2012}. Along these lines, in the panel displaying [NII]$\lambda6583$/[OII]$\lambda3727$  vs. ${\rm M}_*$ (Figure~\ref{fig:n2plots}, right), we also plot low-mass galaxies from the Local Volume Legacy (LVL) Survey \citep{berg2012} that span the mass range probed by the $z=5.0-6.5$ CEERS sample. At $\log({\rm M}_*/M_{\odot})\leq 9.0$, and the corresponding $12+\log({\rm O/H})$ values, there is no scaling between N/O and ${\rm M}_*$ in the $z\sim 0$ dwarf galaxies. Given the evolution of the MZR towards lower metallicity at fixed mass as redshift increases \citep[e.g.,][]{steidel2014,kashino2017,sanders2021}, the CEERS $z=5.0-6.5$  sample must represent an even lower average metallicity at the same mass, and fall even more firmly within the primary nitrogen regime. More generally, if the $z>4$ CEERS sample is indeed in the primary nitrogen regime while the $z<3$ samples are not,  this difference at least partially explains why [NII]$\lambda6583$/H$\alpha$ and O3N2 have a flatter dependence on ${\rm M}_*$ in the $z>4$ samples.
However, such an explanation may not apply to the high-mass $z=4.0-5.0$ bin, which is still significantly lower (higher) in [NII]$\lambda6583$/H$\alpha$ (O3N2) than an interpolation of the $z=2.7-4.0$ CEERS sample at the same ${\rm M}_*$.

\subsection{Line Ratios Including $\alpha$ Elements}
\label{sec:results-alpha}
\begin{figure*}[t!]
\centering
\includegraphics[width=1.0\linewidth]{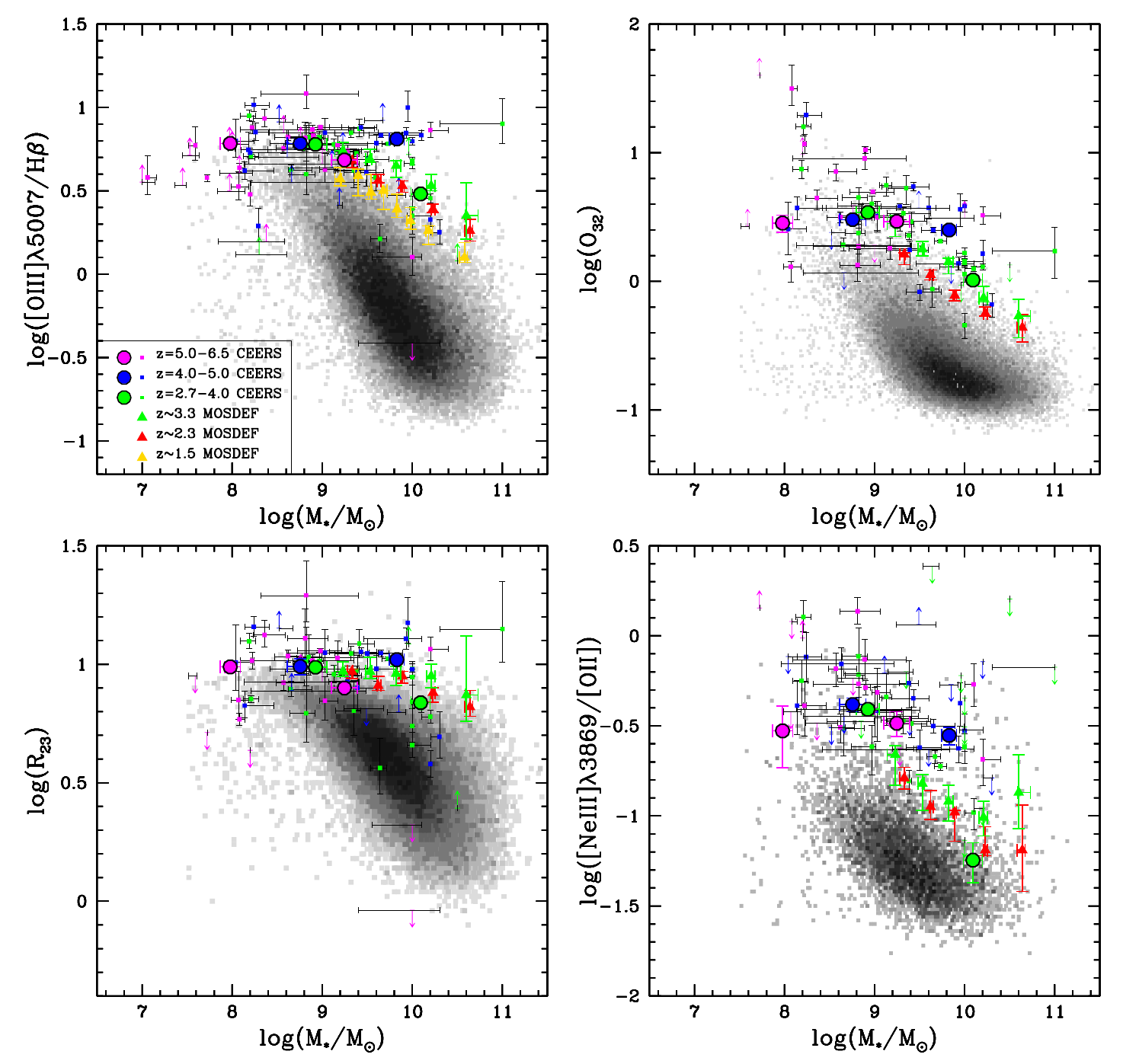}
\caption{$\alpha$-related emission-line ratios vs. stellar mass for CEERS galaxies. These panels shows the line ratios [OIII]$\lambda5007$ (top left), O$_{32}$ (top right), R$_{23}$ (bottom left), and [NeIII]$\lambda3869$/[OII]$\lambda3727$ (bottom right). Symbols are as in Figure~\ref{fig:n2plots}. In addition, measurements from stacked spectra of galaxies in the MOSDEF survey at $z\sim 3.3$ \citep{sanders2021} are shown with large green triangles.
}
\label{fig:alphaplots}
\end{figure*}

We also consider several commonly studied ratios based on $\alpha$ elements (e.g. O, Ne): [OIII]$\lambda5007$/H$\beta$, O$_{32}$, 
R$_{23}$, and [NeIII]$\lambda3869$/[OII]$\lambda3727$. The majority of galaxies in the CEERS $z=2.7-6.5$ sample are detected in [OIII]$\lambda5007$/H$\beta$, O$_{32}$, and R$_{23}$, although individual detections of the fainter [NeIII]$\lambda3869$ line are achieved for only $\sim 40$\% of the sample. As in earlier works \citep[e.g.,][]{juneau2014,coil2015,holden2016}, we find an elevated [OIII]$\lambda5007$/H$\beta$ at fixed stellar mass, relative to the local relation (Figure~\ref{fig:alphaplots}, top left).  We also find a lack of significant evolution in [OIII]$\lambda5007$/H$\beta$ at fixed stellar mass beyond $z\sim 3$. Specifically (with the exception of the high-mass bin at $z=4.0-5.0$), the CEERS emission-line measurements at $z>2.7$ are consistent with those from the MOSDEF $z\sim 3.3$ sample where they overlap in stellar mass. 

While the bulk of $z>1$ measurements shown here follow a trend of increasing [OIII]$\lambda5007$/H$\beta$ as stellar mass decreases, at the lowest masses probed by individual galaxies in the CEERS $z=5.0-6.5$ sample, we find evidence of a turnover towards lower [OIII]$\lambda5007$/H$\beta$. Specifically, several galaxies at the low-mass end of the CEERS sample fall significantly below the SDSS distribution in
[OIII]$\lambda5007$/H$\beta$ vs. ${\rm M}_*$, whereas, above $\log({\rm M}_*/M_{\odot})\sim 9.0$, we find only
one such galaxy. The turnover is also suggested by the trends in the stacked datapoints from both CEERS and MOSDEF.
Together, the CEERS and MOSDEF stacked datapoints trace a monotonically-decreasing sequence in [OIII]$\lambda5007$/H$\beta$ as mass increases above $\log({\rm M}_*/M_{\odot})\sim 9.0$. The CEERS low-mass $z=5.0-6.5$ stack is characterized
by a significantly lower mass ($\log({\rm M}_*/M_{\odot})= 7.98$) and yet its [OIII]$\lambda5007$/H$\beta$ value
is identical to that of the CEERS low-mass stacks at $z=2.7-4.0$ and $z=4.0-5.0$. Such a turnover towards decreasing [OIII]$\lambda5007$/H$\beta$ with decreasing mass at stellar masses below $\log({\rm M}_*/M_{\odot})\sim 9.0$ is expected, given the relationship between [OIII]$\lambda5007$/H$\beta$ and $12+\log({\rm O/H})$ in this regime \citep{curti2020,nakajima2022,sanders2020}. This turnover in [OIII]/H$\beta$ at low stellar mass is also detected by \citet{matthee2022} in {\it JWST}/NIRCam grism observations of a sample of $z\sim 6$ [OIII]-emitting galaxies. 

Like [OIII]$\lambda5007$/H$\beta$, the ratio O$_{32}$ shows no significant evolution at $z\sim 3$ and beyond (Figure~\ref{fig:alphaplots}, top right). O$_{32}$ is most directly sensitive to ionization parameter, and, through the anti-correlation between ionization parameter and oxygen abundance \citep{perezmontero2014}, indirectly traces metallicity \citep{sanders2016,sanders2018}. As in the case of [OIII]$\lambda5007$/H$\beta$, we find that CEERS galaxies at $z=2.7-6.5$ generally have similar O$_{32}$ to those in the $z\sim 3.3$ MOSDEF sample at fixed stellar mass. 

As in the [NII]$\lambda6583$/H$\alpha$ and O3N2 diagrams, the high-mass $z=4.0-5.0$ bin presents as an outlier towards significantly higher [OIII]$\lambda5007$/H$\beta$ and O$_{32}$ at fixed stellar mass relative to the lower-redshift samples (there is no $z=5.0-6.5$ data point in a similar mass range). We require a larger sample, deeper spectroscopy, and robustly-calibrated direct metallicity measurements at $z=4.0-5.0$ \citep{sanders2023b} to determine whether this offset is truly representative of star-forming galaxies at this redshift, and reflective of higher-excitation physical conditions or else individually-undetected AGN activity in some sources.

In the space of R$_{23}$ vs. stellar mass (Figure~\ref{fig:alphaplots}, bottom left), galaxies at $z\geq 2.3$ are offset from the local sequence towards higher R$_{23}$ at fixed stellar mass, yet there is no evolution from $z\sim 2.3$ to $z=6.5$ given that the MOSDEF $z\sim 2.3$ and $z\sim 3.3$ stacks follow the same relation. This behavior differs slightly from what is observed for [OIII]$\lambda5007$/H$\beta$ (and O$_{32}$), where there is slight evolution towards higher line ratio ($\sim 0.1-0.2$~dex) at fixed mass between $z\sim 2.3$ and $z\sim 3.3$, and then no further evolution at higher redshift. 

Finally, we find that [NeIII]$\lambda3869$/[OII]$\lambda3727$ shows evidence of being significantly higher at fixed stellar mass for CEERS galaxies at $z\geq4$ than MOSDEF galaxies at $z\sim 3.3$ (and $z\sim 2.3$). Where we can most robustly gauge this offset, i.e., at $\log({\rm M}_*/M_{\odot})\sim 9.0$ and excluding the CEERS high-mass $z=4.0-5.0$ bin, we find that the the average offset in [NeIII]$\lambda3869$/[OII]$\lambda3727$  between CEERS $z\geq 4.0$ galaxies and MOSDEF $z\sim 3.3$ galaxies is $+0.2$~dex.
At lower masses, there are no average measurements from the MOSDEF survey, so it is not possible to perform a similar comparison. 
We defer further interpretation of the behavior of [NeIII]$\lambda3869$/[OII]$\lambda3727$ at these low masses, as we require larger samples at both $z\geq 4$, and $z\sim2-3$. We do note that, given the close proximity of the constituent lines in the [NeIII]$\lambda3869$/[OII]$\lambda3727$ ratio, the estimate of [NeIII]$\lambda3869$/[OII]$\lambda3727$ is immune to the uncertainties in dust correction and wavelength-dependent flux calibration associated with line ratios widely spaced in wavelength such as O$_{32}$ and R$_{23}$. At the same time, the divergent behavior of O$_{32}$ and [NeIII]$\lambda3869$/[OII]$\lambda3727$ as a function of redshift will require further investigation. Indeed, both ratios are sensitive to the ionization parameter, and both trace $\alpha$ elements \citep{strom2017,witstok2021}, so similar redshift evolution might have been expected.  An elevated ratio of [NeIII]$\lambda3869$ to [OIII]$\lambda5007$ at $z>4$ may be indicative of a harder ionizing spectrum \citep{jeong2020} at fixed nebular metallicity, or higher abundance ratio of neon to oxygen. The former possibility, however, is inconsistent with the conclusions of \citet{sanders2023a} that the ionization conditions do not strongly evolve in CEERS galaxies from $z\sim 2$ to $z\sim 6$. The latter possibility also appears unlikely given the similar enrichment channels of neon and oxygen, and the lack of significant variation in Ne/O detected in local star-forming galaxies as a function of either metallicity or specific SFR \citep{izotov2006}.

\section{Discussion}
\label{sec:discussion}
We have presented a systematic analysis of the empirical relationships between nebular emission-line ratios and stellar mass at $z=2.7-6.5$, based on both individual and stacked composite {\it JWST}/NIRSpec spectra from the CEERS program \citep{finkelstein2022a,finkelstein2022b}.  We analyze common rest-optical line ratios (but some observed for the very first time at $z\geq 3$) based on nitrogen, $\alpha$ elements (oxygen, neon), and hydrogen. Though [NII]$\lambda6583$ proves challenging to detect in individual galaxies at $z\geq 3$, we find based on stacked spectra that there is no strong evolution in the scaling of [NII]$\lambda6583$/H$\alpha$, O3N2, or [NII]$\lambda6583$/[OII]$\lambda3727$ with stellar mass, relative to measurements from the MOSDEF survey at $z<3$ in an overlapping stellar mass range. At the lowest masses covered by the CEERS $z=4.0-5.0$ and $z=5.0-6.5$ samples, we furthermore find evidence for probing a regime of primary nitrogen production. In terms of the $\alpha$ elements, we find no strong evolution at fixed mass beyond $z\sim 3$, in line ratios based on [OII]$\lambda3727$ and [OIII]$\lambda5007$. The empirical nebular emission-line ratios presented here are commonly translated into gas-phase metallicities in order to construct galaxy metallicity scaling relations \citep[e.g.,][]{tremonti2004,sanders2021}, and yet care must be taken when doing so. Specifically, robust calibrations are required for the translation between emission-line ratio and metallicity, and the calibrations adopted for local star-forming galaxies need to be updated for distant galaxies \citep[e.g.,][]{bian2018,sanders2023b}.
We defer the actual translation between line ratio and metallicity to future work, but review here some recent attempts to use {\it JWST} to infer distant galaxy chemical abundances, and the reasons why such analyses are so important for constraining models of galaxy formation.

In addition to detailed investigations of remarkable individual NIRSpec spectra at high redshift
\citep[e.g.,][]{curti2022,williams2022,bunker2023}, {\it JWST} spectra are also starting to be used
for analysis of the chemical abundances of samples of distant galaxies.
\citet{matthee2022} present NIRCam grism observations of [OIII]/H$\beta$ in concert with stellar mass
estimates for a sample of 117 [OIII]-emitting galaxies at $5.3\leq z \leq 6.9$. Metallicities
are estimated from stacked spectra of galaxies in four bins of stellar mass using a single metallicity indicator,
the [OIII]/H$\beta$ ratio, because of the limited wavelength coverage of the NIRCam grism (i.e., $3-4\mu$m).
It is worth noting both that there is very little variation in average [OIII]/H$\beta$ across the four stellar mass bins,
given that the galaxies in the \citet{matthee2022} sample probe the peak of the [OIII]/H$\beta$ distribution,
and, further, that the metallicities inferred from [OIII]/H$\beta$ are significantly higher at fixed mass
than that inferred from a composite spectrum of $z=6.25-6.90$ [OIII] emitters in which the auroral [OIII]$\lambda4363$
line is detected. It will be important to estimate metallicities for this redshift range based on
multiple emission-line ratios like those featured in the current work, including those that vary monotonically with
metallicity (e.g., O$_{32}$, [NeIII]$\lambda3869$/[OII]$\lambda3727$, O3N2), and using calibrations based on direct metallicities at high redshift.

More directly related to the current work, \citet{nakajima2023} attempt to constrain the MZR at $z\sim 4-9$ with {\it JWST}/NIRSpec observations of a sample of 135 galaxies drawn from the publicly available CEERS, GLASS, and ERO programs. Gas-phase metallicities are primarily based on  R$_{23}$ (78 galaxies) or [OIII]$\lambda5007$/H$\beta$ (49 galaxies), with an additional 8 galaxies having direct $T_e$-based metallicities.  Consistent with our empirical results, \citet{nakajima2023} find only weak evolution in the MZR between $z\sim 2-3$ and $z\sim 4-9$. However, we caution that \citet{nakajima2023} have SED-based stellar mass estimates for 81 galaxies in their {\it JWST}  metallicity sample. The remaining 54 galaxies have stellar mass estimates (31 galaxies) or upper limits (23 galaxies) inferred from UV luminosities alone, which carry significant uncertainties due to variations in the mass-to-light ratio. Furthermore, additional work is needed to establish robust metallicity calibrations at $z>3$ based on a larger sample of direct $T_e$-based metallicities at high redshift.

With proper calibration from direct metallicities \citep[e.g.,][]{curti2022,sanders2023a},
{\it JWST} promises to provide excellent observational constraints on the evolution
of the MZR at $z>3$. With a large enough sample and robust SFR estimates, it will also be possible to determine
the nature of the FMR among oxygen abundance, ${\rm M}_*$, and SFR, and whether
it remains as invariant at higher redshift as it does from $z\sim 0$ to $z\sim 3.3$ \citep{sanders2021}. In order to determine the evolution of the MZR, which is meant to represent the average properties of star-forming galaxies at different cosmic epochs, it will be necessary to assemble not only large but also {\it representative} galaxy samples. As discussed in \citet{shapley2023}, the CEERS galaxies we analyze here appear representative of the star-forming main sequence at $z=2.7-5.0$, but with higher-than-average sSFR at $z>5.0$. Given the connections among oxygen abundance, ${\rm M}_*$, and SFR, we also require star-forming galaxy samples at $z>5.0$ that are representative in SFR at fixed ${\rm M}_*$, if we aim to trace MZR evolution at $z>5.0$.

Observational MZR constraints can be compared with theoretical predictions for the evolution of the MZR at the
earliest times. For example, predictions in the literature for this evolution vary considerably. Both
\citet{torrey2019}, based on the IllustrisTNG simulations, and \citet{ma2016}, based on the FIRE simulations,
predict evolution of $\sim -0.3$~dex in gas-phase metallicity at fixed stellar mass from $z=3$ to $z\sim 6$ as galaxy
gas fractions increase, though the normalization of the MZR is significantly higher in \citet{torrey2019}. 
Furthermore, \citet{torrey2019}
predict a linear decline in 12+$\log({\rm O/H})$ with increasing redshift, whereas \citet{ma2016} predict
a flattening in the evolution of the MZR with increasing redshift. In contrast, at slightly
higher redshift ($z=5-8$) and based on the FirstLight simulations, \citet{langan2020} 
predict weak evolution in the normalization of the MZR. The evolution detected in this model is actually
towards {\it increasing} metallicity with increasing redshift, since galaxy gas reservoirs are described
to build up with time (decreasing redshift) over this interval, as gas accretion outpaces star formation and outflows. Furthermore, while the predicted MZR evolution
from \citet{ma2016} is also very shallow within this higher redshift range, the MZR normalization from FIRE is $\sim 0.3$~dex
lower than in the FirstLight simulations. 

Differences among galaxy formation model predictions arise 
because of how the processes of gas inflow, build-up, and outflow are described \citep{langan2020}. Accordingly,
robust measurements of the evolution of the MZR with {\it JWST} will be able to distinguish among the different
prescriptions for baryon cycling at early times. In its first months, {\it JWST} has demonstrated that it is  capable of returning the necessary data to trace the MZR at $z>3$. We now need to assemble sufficient sample sizes (a factor of several larger than the CEERS/NIRSpec $z>3$ sample) with both strong emission-line and stellar mass measurements, and an adequate direct $T_e$-based-metallicity calibration sample for translating emission-line ratios into metallicities.

\section*{Acknowledgements}
We acknowledge the entire CEERS team for their effort to design and execute this Early Release Science
observational program, especially the work to design the MSA observations.
This work is based on observations made with the NASA/
ESA/CSA James Webb Space Telescope. The data were
obtained from the Mikulski Archive for Space Telescopes at
the Space Telescope Science Institute, which is operated by the
Association of Universities for Research in Astronomy, Inc.,
under NASA contract NAS5-03127 for JWST. The specific observations analyzed can be accessed via \dataset[DOI: 10.17909/z7p0-8481]{https://archive.stsci.edu/doi/resolve/resolve.html?doi=10.17909/z7p0-8481}.
We also acknowledge support from NASA grant JWST-GO-01914. Support for this work was also provided through the NASA Hubble Fellowship
grant \#HST-HF2-51469.001-A awarded by the Space Telescope
Science Institute, which is operated by the Association of Universities
for Research in Astronomy, Incorporated, under NASA contract NAS5-26555.


\begin{thebibliography}{}
\expandafter\ifx\csname natexlab\endcsname\relax\def\natexlab#1{#1}\fi

\bibitem[{{Abazajian} {et~al.}(2009){Abazajian}, {Adelman-McCarthy},
  {Ag{\"u}eros}, {Allam}, {Allende Prieto}, {An}, {Anderson}, {Anderson},
  {Annis}, \& {Bahcall}}]{abazajian2009}
{Abazajian}, K.~N., {Adelman-McCarthy}, J.~K., {Ag{\"u}eros}, M.~A., {et~al.}
  2009, \apjs, 182, 543

\bibitem[{{Andrews} \& {Martini}(2013)}]{andrews2013}
{Andrews}, B.~H., \& {Martini}, P. 2013, \apj, 765, 140

\bibitem[{{Arellano-C{\'o}rdova} {et~al.}(2022){Arellano-C{\'o}rdova}, {Berg},
  {Chisholm}, {Arrabal Haro}, {Dickinson}, {Finkelstein}, {Leclercq}, {Rogers},
  {Simons}, {Skillman}, {Trump}, \& {Kartaltepe}}]{arellano2022}
{Arellano-C{\'o}rdova}, K.~Z., {Berg}, D.~A., {Chisholm}, J., {et~al.} 2022,
  \apjl, 940, L23

\bibitem[{{Berg} {et~al.}(2012){Berg}, {Skillman}, {Marble}, {van Zee},
  {Engelbracht}, {Lee}, {Kennicutt}, {Calzetti}, {Dale}, \&
  {Johnson}}]{berg2012}
{Berg}, D.~A., {Skillman}, E.~D., {Marble}, A.~R., {et~al.} 2012, \apj, 754, 98

\bibitem[{{Bian} {et~al.}(2018){Bian}, {Kewley}, \& {Dopita}}]{bian2018}
{Bian}, F., {Kewley}, L.~J., \& {Dopita}, M.~A. 2018, \apj, 859, 175

\bibitem[{{Bunker} {et~al.}(2023){Bunker}, {Saxena}, {Cameron}, {Willott},
  {Curtis-Lake}, {Jakobsen}, {Carniani}, {Smit}, {Maiolino}, {Witstok},
  {Curti}, {D'Eugenio}, {Jones}, {Ferruit}, {Arribas}, {Charlot}, {Chevallard},
  {Giardino}, {de Graaff}, {Looser}, {Luetzgendorf}, {Maseda}, {Rawle}, {Rix},
  {Rodriguez Del Pino}, {Alberts}, {Egami}, {Eisenstein}, {Endsley},
  {Hainline}, {Hausen}, {Johnson}, {Rieke}, {Rieke}, {Robertson}, {Shivaei},
  {Stark}, {Sun}, {Tacchella}, {Tang}, {Williams}, {Willmer}, {Baker}, {Baum},
  {Bhatawdekar}, {Bowler}, {Boyett}, {Chen}, {Circosta}, {Helton}, {Ji}, {Lyu},
  {Nelson}, {Parlanti}, {Perna}, {Sandles}, {Scholtz}, {Suess}, {Topping},
  {Uebler}, {Wallace}, \& {Whitler}}]{bunker2023}
{Bunker}, A.~J., {Saxena}, A., {Cameron}, A.~J., {et~al.} 2023, arXiv e-prints,
  arXiv:2302.07256

\bibitem[{{Cameron} {et~al.}(2023){Cameron}, {Saxena}, {Bunker}, {D'Eugenio},
  {Carniani}, {Maiolino}, {Curtis-Lake}, {Ferruit}, {Jakobsen}, {Arribas},
  {Bonaventura}, {Charlot}, {Chevallard}, {Curti}, {Looser}, {Maseda}, {Rawle},
  {Rodr{\'\i}guez Del Pino}, {Smit}, {{\"U}bler}, {Willott}, {Witstok},
  {Egami}, {Eisenstein}, {Johnson}, {Hainline}, {Rieke}, {Robertson}, {Stark},
  {Tacchella}, {Williams}, {Bhatawdekar}, {Bowler}, {Boyett}, {Circosta},
  {Helton}, {Jones}, {Kumari}, {Ji}, {Nelson}, {Parlanti}, {Sandles},
  {Scholtz}, \& {Sun}}]{cameron2023}
{Cameron}, A.~J., {Saxena}, A., {Bunker}, A.~J., {et~al.} 2023, arXiv e-prints,
  arXiv:2302.04298

\bibitem[{{Cardelli} {et~al.}(1989){Cardelli}, {Clayton}, \&
  {Mathis}}]{cardelli1989}
{Cardelli}, J.~A., {Clayton}, G.~C., \& {Mathis}, J.~S. 1989, \apj, 345, 245

\bibitem[{{Chabrier}(2003)}]{chabrier2003}
{Chabrier}, G. 2003, \pasp, 115, 763

\bibitem[{{Coil} {et~al.}(2015){Coil}, {Aird}, {Reddy}, {Shapley}, {Kriek},
  {Siana}, {Mobasher}, {Freeman}, {Price}, \& {Shivaei}}]{coil2015}
{Coil}, A.~L., {Aird}, J., {Reddy}, N., {et~al.} 2015, \apj, 801, 35

\bibitem[{{Conroy} {et~al.}(2009){Conroy}, {Gunn}, \& {White}}]{conroy2009}
{Conroy}, C., {Gunn}, J.~E., \& {White}, M. 2009, \apj, 699, 486

\bibitem[{{Curti} {et~al.}(2020){Curti}, {Mannucci}, {Cresci}, \&
  {Maiolino}}]{curti2020}
{Curti}, M., {Mannucci}, F., {Cresci}, G., \& {Maiolino}, R. 2020, \mnras, 491,
  944

\bibitem[{{Curti} {et~al.}(2023){Curti}, {D'Eugenio}, {Carniani},
  {Maiolino}, {Sandles}, {Witstok}, {Baker}, {Bennett}, {Piotrowska},
  {Tacchella}, {Charlot}, {Nakajima}, {Maheson}, {Mannucci}, {Amiri},
  {Arribas}, {Belfiore}, {Bonaventura}, {Bunker}, {Chevallard}, {Cresci},
  {Curtis-Lake}, {Hayden-Pawson}, {Jones}, {Kumari}, {Laseter}, {Looser},
  {Marconi}, {Maseda}, {Scholtz}, {Smit}, {{\"U}bler}, \&
  {Wallace}}]{curti2022}
{Curti}, M., {D'Eugenio}, F., {Carniani}, S., {et~al.} 2023, \mnras, 518, 425

\bibitem[{{Ellison} {et~al.}(2008){Ellison}, {Patton}, {Simard}, \&
  {McConnachie}}]{ellison2008}
{Ellison}, S.~L., {Patton}, D.~R., {Simard}, L., \& {McConnachie}, A.~W. 2008,
  \apjl, 672, L107

\bibitem[{{Finkelstein} {et~al.}(2022){Finkelstein}, {Bagley}, {Haro},
  {Dickinson}, {Ferguson}, {Kartaltepe}, {Papovich}, {Burgarella}, {Kocevski},
  {Huertas-Company}, {Iyer}, {Koekemoer}, {Larson}, {P{\'e}rez-Gonz{\'a}lez},
  {Rose}, {Tacchella}, {Wilkins}, {Chworowsky}, {Medrano}, {Morales},
  {Somerville}, {Yung}, {Fontana}, {Giavalisco}, {Grazian}, {Grogin}, {Kewley},
  {Kirkpatrick}, {Kurczynski}, {Lotz}, {Pentericci}, {Pirzkal}, {Ravindranath},
  {Ryan}, {Trump}, {Yang}, {Almaini}, {Amor{\'\i}n}, {Annunziatella},
  {Backhaus}, {Barro}, {Behroozi}, {Bell}, {Bhatawdekar}, {Bisigello}, {Bromm},
  {Buat}, {Buitrago}, {Calabr{\`o}}, {Casey}, {Castellano}, {Ch{\'a}vez Ortiz},
  {Ciesla}, {Cleri}, {Cohen}, {Cole}, {Cooke}, {Cooper}, {Cooray}, {Costantin},
  {Cox}, {Croton}, {Daddi}, {Dav{\'e}}, {de La Vega}, {Dekel}, {Elbaz},
  {Estrada-Carpenter}, {Faber}, {Fern{\'a}ndez}, {Finkelstein}, {Freundlich},
  {Fujimoto}, {Garc{\'\i}a-Argum{\'a}nez}, {Gardner}, {Gawiser},
  {G{\'o}mez-Guijarro}, {Guo}, {Hamblin}, {Hamilton}, {Hathi}, {Holwerda},
  {Hirschmann}, {Hutchison}, {Jaskot}, {Jha}, {Jogee}, {Juneau}, {Jung},
  {Kassin}, {Le Bail}, {Leung}, {Lucas}, {Magnelli}, {Mantha}, {Matharu},
  {McGrath}, {McIntosh}, {Merlin}, {Mobasher}, {Newman}, {Nicholls}, {Pandya},
  {Rafelski}, {Ronayne}, {Santini}, {Seill{\'e}}, {Shah}, {Shen}, {Simons},
  {Snyder}, {Stanway}, {Straughn}, {Teplitz}, {Vanderhoof}, {Vega-Ferrero},
  {Wang}, {Weiner}, {Willmer}, {Wuyts}, {Zavala}, \& {CEERS
  Team}}]{finkelstein2022b}
{Finkelstein}, S.~L., {Bagley}, M.~B., {Haro}, P.~A., {et~al.} 2022, \apjl,
  940, L55

\bibitem[{{Finkelstein} {et~al.}(2023){Finkelstein}, {Bagley}, {Ferguson},
  {Wilkins}, {Kartaltepe}, {Papovich}, {Yung}, {Arrabal Haro}, {Behroozi},
  {Dickinson}, {Kocevski}, {Koekemoer}, {Larson}, {Le Bail}, {Morales},
  {P{\'e}rez-Gonz{\'a}lez}, {Burgarella}, {Dav{\'e}}, {Hirschmann},
  {Somerville}, {Wuyts}, {Bromm}, {Casey}, {Fontana}, {Fujimoto}, {Gardner},
  {Giavalisco}, {Grazian}, {Grogin}, {Hathi}, {Hutchison}, {Jha}, {Jogee},
  {Kewley}, {Kirkpatrick}, {Long}, {Lotz}, {Pentericci}, {Pierel}, {Pirzkal},
  {Ravindranath}, {Ryan}, {Trump}, {Yang}, {Bhatawdekar}, {Bisigello}, {Buat},
  {Calabr{\`o}}, {Castellano}, {Cleri}, {Cooper}, {Croton}, {Daddi}, {Dekel},
  {Elbaz}, {Franco}, {Gawiser}, {Holwerda}, {Huertas-Company}, {Jaskot},
  {Leung}, {Lucas}, {Mobasher}, {Pandya}, {Tacchella}, {Weiner}, \&
  {Zavala}}]{finkelstein2022a}
{Finkelstein}, S.~L., {Bagley}, M.~B., {Ferguson}, H.~C., {et~al.} 2023, \apjl,
  946, L13

\bibitem[{{Fujimoto} {et~al.}(2023){Fujimoto}, {Arrabal Haro}, {Dickinson},
  {Finkelstein}, {Kartaltepe}, {Larson}, {Burgarella}, {Bagley}, {Behroozi},
  {Chworowsky}, {Hirschmann}, {Trump}, {Wilkins}, {Yung}, {Koekemoer},
  {Papovich}, {Pirzkal}, {Ferguson}, {Fontana}, {Grogin}, {Grazian}, {Kewley},
  {Kocevski}, {Lotz}, {Pentericci}, {Ravindranath}, {Somerville}, {Amorin},
  {Backhaus}, {Calabro}, {Casey}, {Cooper}, {Franco}, {Giavalisco}, {Hathi},
  {Harish}, {Hutchison}, {Iyer}, {Jung}, {Lucas}, \& {Zavala}}]{fujimoto2023}
{Fujimoto}, S., {Arrabal Haro}, P., {Dickinson}, M., {et~al.} 2023, arXiv
  e-prints, arXiv:2301.09482

\bibitem[{{Holden} {et~al.}(2016){Holden}, {Oesch}, {Gonz{\'a}lez},
  {Illingworth}, {Labb{\'e}}, {Bouwens}, {Franx}, {van Dokkum}, \&
  {Spitler}}]{holden2016}
{Holden}, B.~P., {Oesch}, P.~A., {Gonz{\'a}lez}, V.~G., {et~al.} 2016, \apj,
  820, 73

\bibitem[{{Izotov} {et~al.}(2006){Izotov}, {Stasi{\'n}ska}, {Meynet}, {Guseva},
  \& {Thuan}}]{izotov2006}
{Izotov}, Y.~I., {Stasi{\'n}ska}, G., {Meynet}, G., {Guseva}, N.~G., \&
  {Thuan}, T.~X. 2006, \aap, 448, 955

\bibitem[{{Jeong} {et~al.}(2020){Jeong}, {Shapley}, {Sanders}, {Runco},
  {Topping}, {Reddy}, {Kriek}, {Coil}, {Mobasher}, {Siana}, {Shivaei},
  {Freeman}, {Azadi}, {Price}, {Leung}, {Fetherolf}, {de Groot}, {Zick},
  {Fornasini}, \& {Barro}}]{jeong2020}
{Jeong}, M.-S., {Shapley}, A.~E., {Sanders}, R.~L., {et~al.} 2020, \apjl, 902,
  L16

\bibitem[{{Juneau} {et~al.}(2014){Juneau}, {Bournaud}, {Charlot}, {Daddi},
  {Elbaz}, {Trump}, {Brinchmann}, {Dickinson}, {Duc}, {Gobat}, {Jean-Baptiste},
  {Le Floc'h}, {Lehnert}, {Pacifici}, {Pannella}, \& {Schreiber}}]{juneau2014}
{Juneau}, S., {Bournaud}, F., {Charlot}, S., {et~al.} 2014, \apj, 788, 88

\bibitem[{{Kashino} {et~al.}(2017){Kashino}, {Silverman}, {Sanders},
  {Kartaltepe}, {Daddi}, {Renzini}, {Valentino}, {Rodighiero}, {Juneau},
  {Kewley}, {Zahid}, {Arimoto}, {Nagao}, {Chu}, {Sugiyama}, {Civano}, {Ilbert},
  {Kajisawa}, {Le F{\`e}vre}, {Maier}, {Masters}, {Miyaji}, {Onodera},
  {Puglisi}, \& {Taniguchi}}]{kashino2017}
{Kashino}, D., {Silverman}, J.~D., {Sanders}, D., {et~al.} 2017, \apj, 835, 88

\bibitem[{{Kriek} {et~al.}(2009){Kriek}, {van Dokkum}, {Labb{\'e}}, {Franx},
  {Illingworth}, {Marchesini}, \& {Quadri}}]{kriek2009}
{Kriek}, M., {van Dokkum}, P.~G., {Labb{\'e}}, I., {et~al.} 2009, \apj, 700,
  221

\bibitem[{{Langan} {et~al.}(2020){Langan}, {Ceverino}, \&
  {Finlator}}]{langan2020}
{Langan}, I., {Ceverino}, D., \& {Finlator}, K. 2020, \mnras, 494, 1988

\bibitem[{{Langeroodi} {et~al.}(2022){Langeroodi}, {Hjorth}, {Chen}, {Kelly},
  {Williams}, {Lin}, {Scarlata}, {Zitrin}, {Broadhurst}, {Diego}, {Huang},
  {Filippenko}, {Foley}, {Jha}, {Koekemoer}, {Oguri}, {Perez-Fournon},
  {Pierel}, {Poidevin}, \& {Strolger}}]{langeroodi2022}
{Langeroodi}, D., {Hjorth}, J., {Chen}, W., {et~al.} 2022, arXiv e-prints,
  arXiv:2212.02491

\bibitem[{{Ma} {et~al.}(2016){Ma}, {Hopkins}, {Faucher-Gigu{\`e}re}, {Zolman},
  {Muratov}, {Kere{\v{s}}}, \& {Quataert}}]{ma2016}
{Ma}, X., {Hopkins}, P.~F., {Faucher-Gigu{\`e}re}, C.-A., {et~al.} 2016,
  \mnras, 456, 2140

\bibitem[{{Mannucci} {et~al.}(2010){Mannucci}, {Cresci}, {Maiolino}, {Marconi},
  \& {Gnerucci}}]{mannucci2010}
{Mannucci}, F., {Cresci}, G., {Maiolino}, R., {Marconi}, A., \& {Gnerucci}, A.
  2010, \mnras, 408, 2115

\bibitem[{{Mascia} {et~al.}(2023){Mascia}, {Pentericci}, {Calabr{\`o}}, {Treu},
  {Santini}, {Yang}, {Napolitano}, {Roberts-Borsani}, {Bergamini}, {Grillo},
  {Rosati}, {Vulcani}, {Castellano}, {Boyett}, {Fontana}, {Glazebrook},
  {Henry}, {Mason}, {Merlin}, {Morishita}, {Nanayakkara}, {Paris}, {Roy},
  {Williams}, {Wang}, {Brammer}, {Brada{\v{c}}}, {Chen}, {Kelly}, {Koekemoer},
  {Trenti}, \& {Windhorst}}]{mascia2023}
{Mascia}, S., {Pentericci}, L., {Calabr{\`o}}, A., {et~al.} 2023, \aap, 672,
  A155

\bibitem[{{Matthee} {et~al.}(2022){Matthee}, {Mackenzie}, {Simcoe}, {Kashino},
  {Lilly}, {Bordoloi}, \& {Eilers}}]{matthee2022}
{Matthee}, J., {Mackenzie}, R., {Simcoe}, R.~A., {et~al.} 2022, arXiv e-prints,
  arXiv:2211.08255

\bibitem[{{Nakajima} {et~al.}(2023){Nakajima}, {Ouchi}, {Isobe}, {Harikane},
  {Zhang}, {Ono}, {Umeda}, \& {Oguri}}]{nakajima2023}
{Nakajima}, K., {Ouchi}, M., {Isobe}, Y., {et~al.} 2023, arXiv e-prints,
  arXiv:2301.12825

\bibitem[{{Nakajima} {et~al.}(2022){Nakajima}, {Ouchi}, {Xu}, {Rauch},
  {Harikane}, {Nishigaki}, {Isobe}, {Kusakabe}, {Nagao}, {Ono}, {Onodera},
  {Sugahara}, {Kim}, {Komiyama}, {Lee}, \& {Zahedy}}]{nakajima2022}
{Nakajima}, K., {Ouchi}, M., {Xu}, Y., {et~al.} 2022, \apjs, 262, 3

\bibitem[{{Onodera} {et~al.}(2016){Onodera}, {Carollo}, {Lilly}, {Renzini},
  {Arimoto}, {Capak}, {Daddi}, {Scoville}, {Tacchella}, {Tatehora}, \&
  {Zamorani}}]{onodera2016}
{Onodera}, M., {Carollo}, C.~M., {Lilly}, S., {et~al.} 2016, \apj, 822, 42

\bibitem[{{P{\'e}rez-Montero}(2014)}]{perezmontero2014}
{P{\'e}rez-Montero}, E. 2014, \mnras, 441, 2663

\bibitem[{{Pilyugin} {et~al.}(2012){Pilyugin}, {Grebel}, \&
  {Mattsson}}]{pilyugin2012}
{Pilyugin}, L.~S., {Grebel}, E.~K., \& {Mattsson}, L. 2012, \mnras, 424, 2316

\bibitem[{{Reddy} {et~al.}(2023){Reddy}, {Topping}, {Sanders}, {Shapley}, \&
  {Brammer}}]{reddy2023}
{Reddy}, N.~A., {Topping}, M.~W., {Sanders}, R.~L., {Shapley}, A.~E., \&
  {Brammer}, G. 2023, arXiv e-prints, arXiv:2301.07249

\bibitem[{{Reddy} {et~al.}(2020){Reddy}, {Shapley}, {Kriek}, {Steidel},
  {Shivaei}, {Sanders}, {Mobasher}, {Coil}, {Siana}, {Freeman}, {Azadi},
  {Fetherolf}, {Leung}, {Price}, \& {Zick}}]{reddy2020}
{Reddy}, N.~A., {Shapley}, A.~E., {Kriek}, M., {et~al.} 2020, \apj, 902, 123

\bibitem[{{Sanders} {et~al.}(2023{\natexlab{a}}){Sanders}, {Shapley},
  {Topping}, {Reddy}, \& {Brammer}}]{sanders2023b}
{Sanders}, R.~L., {Shapley}, A.~E., {Topping}, M.~W., {Reddy}, N.~A., \&
  {Brammer}, G.~B. 2023{\natexlab{a}}, arXiv e-prints, arXiv:2303.08149

\bibitem[{{Sanders} {et~al.}(2023{\natexlab{b}}){Sanders}, {Shapley},
  {Topping}, {Reddy}, \& {Brammer}}]{sanders2023a}
---. 2023{\natexlab{b}}, arXiv e-prints, arXiv:2301.06696

\bibitem[{{Sanders} {et~al.}(2016){Sanders}, {Shapley}, {Kriek}, {Reddy},
  {Freeman}, {Coil}, {Siana}, {Mobasher}, {Shivaei}, {Price}, \& {de
  Groot}}]{sanders2016}
{Sanders}, R.~L., {Shapley}, A.~E., {Kriek}, M., {et~al.} 2016, \apj, 816, 23

\bibitem[{{Sanders} {et~al.}(2018){Sanders}, {Shapley}, {Kriek}, {Freeman},
  {Reddy}, {Siana}, {Coil}, {Mobasher}, {Dav{\'e}}, {Shivaei}, {Azadi},
  {Price}, {Leung}, {Fetherolf}, {de Groot}, {Zick}, {Fornasini}, \&
  {Barro}}]{sanders2018}
---. 2018, \apj, 858, 99

\bibitem[{{Sanders} {et~al.}(2020){Sanders}, {Shapley}, {Reddy}, {Kriek},
  {Siana}, {Coil}, {Mobasher}, {Shivaei}, {Freeman}, {Azadi}, {Price}, {Leung},
  {Fetherolf}, {de Groot}, {Zick}, {Fornasini}, \& {Barro}}]{sanders2020}
{Sanders}, R.~L., {Shapley}, A.~E., {Reddy}, N.~A., {et~al.} 2020, \mnras, 491,
  1427

\bibitem[{{Sanders} {et~al.}(2021){Sanders}, {Shapley}, {Jones}, {Reddy},
  {Kriek}, {Siana}, {Coil}, {Mobasher}, {Shivaei}, {Dav{\'e}}, {Azadi},
  {Price}, {Leung}, {Freeman}, {Fetherolf}, {de Groot}, {Zick}, \&
  {Barro}}]{sanders2021}
{Sanders}, R.~L., {Shapley}, A.~E., {Jones}, T., {et~al.} 2021, \apj, 914, 19

\bibitem[{{Shapley} {et~al.}(2023){Shapley}, {Sanders}, {Reddy}, {Topping}, \&
  {Brammer}}]{shapley2023}
{Shapley}, A.~E., {Sanders}, R.~L., {Reddy}, N.~A., {Topping}, M.~W., \&
  {Brammer}, G.~B. 2023, arXiv e-prints, arXiv:2301.03241

\bibitem[{{Speagle} {et~al.}(2014){Speagle}, {Steinhardt}, {Capak}, \&
  {Silverman}}]{speagle2014}
{Speagle}, J.~S., {Steinhardt}, C.~L., {Capak}, P.~L., \& {Silverman}, J.~D.
  2014, \apjs, 214, 15

\bibitem[{{Steidel} {et~al.}(2014){Steidel}, {Rudie}, {Strom}, {Pettini},
  {Reddy}, {Shapley}, {Trainor}, {Erb}, {Turner}, {Konidaris}, {Kulas}, {Mace},
  {Matthews}, \& {McLean}}]{steidel2014}
{Steidel}, C.~C., {Rudie}, G.~C., {Strom}, A.~L., {et~al.} 2014, \apj, 795, 165

\bibitem[{{Strom} {et~al.}(2017){Strom}, {Steidel}, {Rudie}, {Trainor},
  {Pettini}, \& {Reddy}}]{strom2017}
{Strom}, A.~L., {Steidel}, C.~C., {Rudie}, G.~C., {et~al.} 2017, \apj, 836, 164

\bibitem[{{Tang} {et~al.}(2023){Tang}, {Stark}, {Chen}, {Mason}, {Topping},
  {Endsley}, {Senchyna}, {Plat}, {Lu}, {Whitler}, {Robertson}, \&
  {Charlot}}]{tang2023}
{Tang}, M., {Stark}, D.~P., {Chen}, Z., {et~al.} 2023, arXiv e-prints,
  arXiv:2301.07072

\bibitem[{{Topping} {et~al.}(2021){Topping}, {Shapley}, {Sanders}, {Kriek},
  {Reddy}, {Coil}, {Mobasher}, {Siana}, {Freeman}, {Shivaei}, {Azadi}, {Price},
  {Leung}, {Fetherolf}, {de Groot}, {Zick}, {Fornasini}, {Barro}, \&
  {Runco}}]{topping2021}
{Topping}, M.~W., {Shapley}, A.~E., {Sanders}, R.~L., {et~al.} 2021, \mnras,
  506, 1237

\bibitem[{{Torrey} {et~al.}(2019){Torrey}, {Vogelsberger}, {Marinacci},
  {Pakmor}, {Springel}, {Nelson}, {Naiman}, {Pillepich}, {Genel}, {Weinberger},
  \& {Hernquist}}]{torrey2019}
{Torrey}, P., {Vogelsberger}, M., {Marinacci}, F., {et~al.} 2019, \mnras, 484,
  5587

\bibitem[{{Tremonti} {et~al.}(2004){Tremonti}, {Heckman}, {Kauffmann},
  {Brinchmann}, {Charlot}, {White}, {Seibert}, {Peng}, {Schlegel}, {Uomoto},
  {Fukugita}, \& {Brinkmann}}]{tremonti2004}
{Tremonti}, C.~A., {Heckman}, T.~M., {Kauffmann}, G., {et~al.} 2004, \apj, 613,
  898

\bibitem[{{Williams} {et~al.}(2022){Williams}, {Kelly}, {Chen}, {Brammer},
  {Zitrin}, {Treu}, {Scarlata}, {Koekemoer}, {Oguri}, {Lin}, {Diego}, {Nonino},
  {Hjorth}, {Langeroodi}, {Broadhurst}, {Rogers}, {Perez-Fournon}, {Foley},
  {Jha}, {Filippenko}, {Strolger}, {Pierel}, {Poidevin}, \&
  {Yang}}]{williams2022}
{Williams}, H., {Kelly}, P.~L., {Chen}, W., {et~al.} 2022, arXiv e-prints,
  arXiv:2210.15699

\bibitem[{{Witstok} {et~al.}(2021){Witstok}, {Smit}, {Maiolino}, {Curti},
  {Laporte}, {Massey}, {Richard}, \& {Swinbank}}]{witstok2021}
{Witstok}, J., {Smit}, R., {Maiolino}, R., {et~al.} 2021, \mnras, 508, 1686

\end{thebibliography}

\end{document}